\documentclass[journal, onecolumn]{IEEEtran}
\pdfoutput=1
\usepackage[T1]{fontenc}
\usepackage[latin1]{inputenc}
\usepackage{amssymb}
\usepackage[cmex10]{amsmath}
\usepackage[]{graphicx}
\usepackage[caption=false]{subfig}
\usepackage{booktabs}
\usepackage{multirow}
\usepackage{array}
\usepackage{xcolor}
\usepackage{epstopdf}
\usepackage{url}
\usepackage{bm} 
\usepackage{algorithm, algorithmicx, algpseudocode}
\usepackage{float} 
\usepackage{grffile} 
\usepackage{tablefootnote}

\usepackage{eqparbox}

\algtext*{EndWhile}
\algtext*{EndIf}
\algtext*{EndFunction}

\ifCLASSINFOpdf
\else
\fi

\begin{document}

\title{Distributed mining of large scale remote sensing image archives on public computing infrastructures}

\author{Luigi~Mascolo,~
        Marco~Quartulli,~ 
				Pietro~Guccione,~
				Giovanni~Nico~
				and~Igor~G.~Olaizola
\thanks{L.~Mascolo and P.~Guccione are with Department of Electrical and Information Engineering, Polytechnic of Bari, Via E. Orabona 4, 70125 Bari, Italy e-mail: lui.mascolo@gmail.com.}%
\thanks{M.~Quartulli and I.~G.~Olaizola are with Vicomtech--IK4, Mikeletegi Pasealekua 57, 20009 Donostia--San Sebasti\'{a}n, Spain e-mail: \mbox{mquartulli@vicomtech.org.}}%
\thanks{G.~Nico is with Istituto per le Applicazioni del Calcolo, National Research Council, Via Amendola 122, 70126 Bari, Italy.}%
}


\maketitle


\begin{abstract}

Earth Observation (EO) mining aims at supporting efficient access to and exploration of petabyte-scale space- and airborne remote sensing archives that are currently expanding at rates of terabytes per day. A significant challenge is performing the analysis required by envisaged applications --- like for instance process mapping for environmental risk management --- in reasonable time. 
In this work, we address the problem of content-based image retrieval via example-based queries from EO data archives. In particular, we focus on the analysis of polarimetric SAR data, for which target decomposition theorems have proved fundamental in discovering patterns in data and characterize the ground scattering properties. 
To this end, we propose an interactive region-oriented content-based image mining system in which 1) unsupervised ingestion processes are distributed onto virtual machines in elastic, on-demand computing infrastructures 2) archive-scale content hierarchical indexing is implemented in terms of a ``big data" analytics cluster-computing framework 3) query processing amounts to traversing the generated binary tree index, computing distances that correspond to descriptor-based similarity measures between image groups and a query image tile.
We describe in depth both the strategies and the actual implementations for the ingestion and indexing components, and verify the approach by experiments carried out on the NASA/JPL UAVSAR full polarimetric data archive.

We report the results of the tests performed on computer clusters by using a public Infrastructure-as-a-Service and evaluating the impact of  cluster configuration on system performance. Results are promising for data mapping and information retrieval applications.
\end{abstract}

\begin{IEEEkeywords}
Content--Based Retrieval, Remote Sensing, Cloude-Pottier decomposition, Cloud Computing, Big Data.
\end{IEEEkeywords}

\IEEEpeerreviewmaketitle

\thispagestyle{plain} 
\pagestyle{plain} 

\section{Introduction}
\label{sec:introduction}

\IEEEPARstart{R}{emotely} sensed data volumes are growing at faster and faster rates due to the increasing number of spaceborne and airborne Earth Observation (EO) missions and to the improvement of image resolution. As an example, according to a yearly report published at the end of 2013, the archives of NASA's Earth Observing System Data and Information System (EOSDIS) have a volume of almost 10 petabytes, with 6,900 accessible datasets and an average archive growth of approximately 8.5 terabytes per day. The possibility of accessing and processing large databases of remote sensing images allows planetary scale applications like deforestation monitoring, glacial retreat investigation, urban development mapping, land cover classification and so on.
Allowing an efficient discovery, annotation and retrieval of data products is the goal of EO mining systems~\cite{quartulli2013review}.

Common methods to search for images are based on meta-data, that is a user formulates a query in terms of geographic location, sensor parameters, time of acquisition, manual annotations, and the system returns images that are relevant to the query.
The problem with such approaches is that they hardly satisfy user needs in terms of human perception, while manually annotating large volumes of images for describing their content is so expensive and time-consuming that it becomes impossible when facing real scenarios, where petabyte-scale image archives are the rule.
An alternative approach is Content-Based Image Retrieval (CBIR), where a description of an image in terms of primitive features (textures, shapes, relevant colors etc\dots) is automatically extracted from its content, providing the user with query and retrieval methods that are closer to  visual perception.
Despite several systems for content-based retrieval having been proposed for mining EO image databases, most of them have not been conceived for scaling to massive volumes of data and it is estimated that up to 95\% of the data present in existing archives has never been accessed or analyzed by a human analyst~\cite{koubarakis2012building}.

The fundamental limitation of remote sensing content information retrieval has been about their applicability to extended archives: operational scalability has not usually been a characteristic of a number of prototype systems presented in the literature.
A system for automatic object and tile-based feature extraction, content mining and indexing has been proposed in~\cite{shyu2007geoiris,klaric2006framework}. 
Experiments are conducted on a 40 Gigabyte archive of high resolution images from the IKONOS and QuickBird satellites. 
Multiple machines are used to answer user queries, with each machine managing a different feature related index and processing queries in parallel. 
When queried, the system answers by combining the results from the distributed index. 
However, in substance, both the system architecture and the algorithms are single node-based, which is unlikely to provide the computational capabilities required to process millions or billions of records. 
Ingestion and, in particular, indexing procedures are considered offline operations.
In~\cite{moise2013terabyte} a similarity-based retrieval system for Terabyte scale image archives is presented. 
The reported experiments are conducted on a large image dataset whose descriptors are stored on a cluster of commodity machines.
However parallel processing procedures needed for the generation of the descriptors are not considered, and index creation is performed by building a tree-structured index on a single machine, relying on the descriptors of a limited subset of archived products obtained by randomly sampling the entire collection.
Distributing index creation procedures across multiple machines is in fact a challenging task, consisting of long-running data- and computational-intensive operations for which scalable data analysis algorithms are needed.

In this paper, we propose an EO data search framework that allows the exploration of massive datasets via example-based queries. 
We show that the system can be exploited to interactively query large-scale EO imagery archives with logarithmic search complexity and report results of tests  performed  by  using  a  public  Infrastructure-as-a-Service.
Without imposing any constraint on the extension of the system to any kind of sensor, we focus in this specific work on polarimetric Synthetic Aperture Radar (SAR) data, where the improvements in the sensor resolution and the use of multiple polarizations are causing a relevant growth of data volume. 
The publicly available database of polarimetric SAR images collected by the UAVSAR~\cite{rosen2006uavsar} sensor has been considered as a case study.

The paper is structured as follows. In Section~\ref{sec:methodology}, the analysis and parallelization methodologies involved in different parts of the system architecture are introduced. 
In Section~\ref{sec:polarimetric}, the basic concepts of target decomposition theory for polarimetric SAR data are introduced and the specific technique employed in the system, Cloude-Pottier  decomposition, is briefly described.
In Section~\ref{sec:sys_architecture} the system architecture and the procedures adopted for the data ingestion and the index creation are outlined. 
The results of the experiments carried out by using cloud-based computing infrastructures are reported in Section~\ref{sec:experiments}. 
Finally, in Section~\ref{sec:conclusions}, a discussion on the results and the conclusions close the article.


\section{Analysis and parallelization methodology}
\label{sec:methodology}
CBIR systems are typically based on three separate data processing phases and on a separate client UI. 

The first phase entails an unsupervised ingestion step, in which single archive products are analyzed independently of each other in order to characterize them based on their contents. 
The second phase consists of an indexing step in which the obtained characterizations are compared with each other in order to organize the contents of the archive in groups of similar characteristics, in order to simplify navigation. This phase is unsupervised as well, and results in the production of a content index.
A last and interactive phase is query processing, in which the unsupervised characterization of an example query is used to navigate the index tree to retrieve the most relevant archive items.

In general terms, the ingestion phase can be seen as ``perfectly parallel'' or ``embarrassingly parallel''~\cite{foster1995designing} and fits a single instruction, multiple data paradigm, while the second phase, index formation, requires iterative processes in which archive elements are repeatedly compared with each other. The third phase, query processing, already operates on efficient data structures and often does not require specific parallelization methodologies.

\begin{figure}[t]
	\centering
	\subfloat[]
		{\label{fig:mapPattern} \includegraphics[height=4cm]{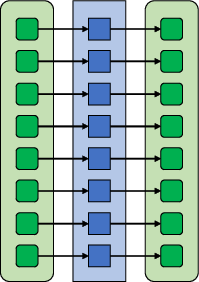}}\qquad
  \subfloat[]
		{\label{fig:redPattern} \includegraphics[height=4cm]{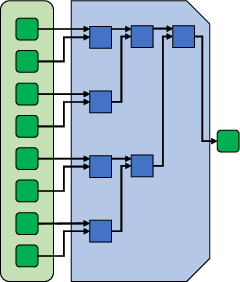}}
	\caption{Concept diagrams: parallel implementations of a map pattern and a reduce pattern. 
	In the map pattern shown in Fig~\ref{fig:mapPattern}, an elemental function is independently applied over every element of a data collection. The reduce pattern performs pairwise combinations of the elements of the collection in order to create a summary value, as shown in Fig~\ref{fig:redPattern}.}
		\label{fig:mapRedPattern}
\end{figure}

Perfectly parallel workloads typically involve multiple tasks that have no dependency on one another. It is in this sense that the first phase can be expressed by a \emph{map} pattern (see Fig.~\ref{fig:mapPattern}) that invokes an elemental function over every element of a data collection in parallel~\cite{mccool2012structured}.
A fundamental requirement for the implemented elemental functions is to not modify global data on which other instances of that function depend.
Examples of operations amenable to direct implementation as map-like functions that are typically part of the ingestion phase of an EO image mining system include data product access, tiling and feature extraction.
All such functions operate on isolated elements of the input archive, but their usage is generally not uniform in time.
Virtual resources in elastic cloud computing infrastructures are well suited to such a scenario, allowing to distribute the ingestion processes dynamically and on-demand, achieving economies of scale and maximizing the effectiveness of the shared resources.

The scenario of the second phase is different: the dataset has to be seen as a whole rather than a collection of separate elements, as the iterative machine-learning processes involved require communication between tasks, especially when combining intermediate results.
The combine function defines a \emph{reduce} pattern, with the elements of a collection being combined in any order to create a summary (see Fig.~\ref{fig:redPattern}).

MapReduce is a programming model that is well suited to this scenario.
It requires the definition of two user-defined methods that process data organized as key/value pairs. The map function takes in input  key/value pairs and, for each of them, applies the mapping procedures, resulting in zero, one or multiple key/value pairs in output. The reduce function takes the intermediate  key/value groups, i.e. the grouped-by-key/value pairs, and processes together the intermediate values associated with the same key. 
A key advantage of such programming model is the possibility of keeping the computation close to the data, avoiding the need of moving the data from the place where they are stored.
Expressing an algorithm by sequences of map and reduce patterns allows implementing parallel applications in a way that is suitable for the execution on large clusters of commodity machines~\cite{dean2008mapreduce}.

Based on these considerations and to demonstrate the feasibility of large scale remote sensing CBIR on inexpensive on-demand processing infrastructures, we have developed a complete system that allows to distribute both ingestion and indexing processes over computer clusters and is able to scale to large scale databases.




\section{Mining polarimetric SAR data archives}
\label{sec:polarimetric}
\begin{figure*}[t]
	\centering
	\includegraphics[width=0.70\columnwidth]{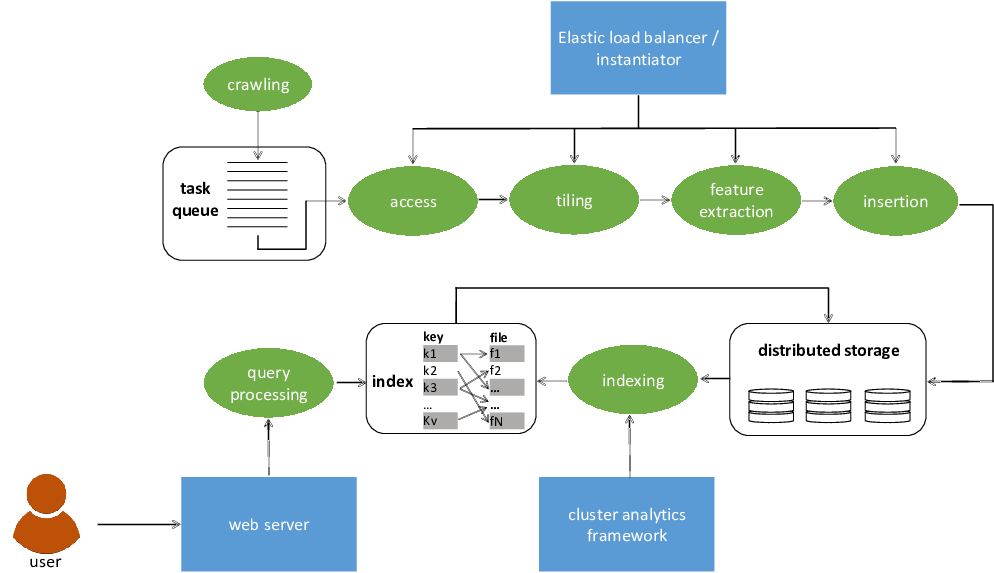}
	\caption{A sketch of the high level architectural decomposition of the system. In the ingestion phase, in the upper part of the diagram, crawlers populate a task queue with identifiers of products to be ingested. One or more workers instantiated by an elastic load balancing mechanism access the data products, divide them into tiles, extract descriptors for each of them (feature extraction) and insert these descriptors into a distributed storage system. In the indexing phase, in the bottom half of the diagram, a cluster of machines is orchestrated by a cluster analytics framework to build a content-based index of the items in the distributed storage database. The results of the indexing  process consist of a data structure  that allows logarithmic-complexity content queries to take place. They are made available to  a query processing system that is invoked via a web server based on interface events generated by a user in an interactive visualization subsystem.}
	\label{fig:highLevArch}
\end{figure*}

The potential of SAR polarimetry to characterize the physical properties of the Earth surface has led to a variety of applications that aim at exploiting the ground scattering mechanism to extract geophysical parameters and perform landcover classification. Such parameter signatures are known to allow ground target classification, and therefore represent a direct possibility for the development of a system to mine large scale remote sensing archives. 
 

Electromagnetic radiation incident on a target with a given polarization (horizontal or vertical) produces a backscattered wave that has, in general, both horizontal and vertical polarizations. The way a target changes the backscattered wave polarization is described by the scattering matrix, which describes the transformation of the electric field between the incident and backscattered wave:
\begin{equation}
	\left[ {\begin{array}{*{20}{c}}
	{E_H^b}\\
	{E_V^b}
	\end{array}} \right] = \left[ {\begin{array}{*{20}{c}}
	{{S_{HH}}}&{{S_{HV}}}\\
	{{S_{VH}}}&{{S_{VV}}}
	\end{array}} \right]\left[ {\begin{array}{*{20}{c}}
	{E_H^i}\\
	{E_V^i}
	\end{array}} \right] \text{,}
\end{equation}
where ${{\bf{E}}^i} = {\left[ {\begin{array}{*{20}{c}} {E_H^i}&{E_V^i} \end{array}} \right]^T}$ is the electric field of the incident wave, ${{\bf{E}}^b} = {\left[ {\begin{array}{*{20}{c}} {E_H^b}&{E_V^b} \end{array}} \right]^T}$ is the electric field of the backscattered wave and ${S_{HH}}$, ${S_{HV}}$, ${S_{VH}}$ and ${S_{VV}}$ are the four complex coefficients of the scattering matrix, which are variable and depend on the nature of the target~\cite{lee2009polarimetric}.

The elements on the main diagonal of the scattering matrix, ${S_{HH}}$ and ${S_{VV}}$ , produce the power return in the copolarized channels and the elements ${S_{HV}}$ and ${S_{VH}}$ produce the power return in the cross-polarized channels. 

There are several ways to represent the scattering properties of a target. The coherence matrix is a possible power representation on which Cloude-Pottier decomposition theory relies~\cite{cloude1996review}. Let us consider the vectorized version of the scattering matrix, based on Pauli spin elements:
\begin{equation}
	{{\bf{k}}_P} = \frac{1}{{\sqrt 2 }}\left[ {\begin{array}{*{20}{c}}
	{{S_{HH}} + {S_{VV}}}\\
	{{S_{HH}} - {S_{VV}}}\\
	{2{S_{HV}}}
	\end{array}} \right] \text{,}
\end{equation}
where the reciprocity assumption ${S_{HV}} = {S_{VH}}$ has been supposed. The coherence matrix is a positive semi-definite Hermitian matrix, given by the product of the Pauli vector by itself:
\begin{equation}
	{\bf{T}} = {{\bf{k}}_P}{\bf{k}}_P^T \text{.}
\end{equation}
Cloude-Pottier decomposition relies on the eigen-decomposition of the (averaged) coherence matrix:
\begin{equation}
	{\bf{T}} = {\bf{U\Sigma }}{{\bf{U}}^{ - 1}} = \sum\limits_{i = 1}^3 {{\lambda _i}{{\bf{u}}_i} \cdot {\bf{u}}_i^{T*}} \text{,}
\end{equation}
where ${\bf{\Sigma }} = diag\left( {{\lambda _1},{\lambda _2},{\lambda _3}} \right)$ is a diagonal matrix with non-negative real components, corresponding to the eigenvalues of ${\bf{T}}$, and ${\bf{U}}$ is a square matrix whose columns are the eigenvectors of ${\bf{T}}$. Let us further account for the following parameterization of the eigenvectors, holding for the case of scattering medium that does not have azimuth symmetry: 
\begin{equation}
{{\bf{u}}_i} = {e^{j{\phi _i}}}\left[ {\begin{array}{*{20}{c}}
{\cos {\alpha _i}}\\
{\sin {\alpha _i}\cos {\beta _i}{e^{j{\delta _i}}}}\\
{\sin {\alpha _i}sin{\beta _i}{e^{j{\gamma _i}}}}
\end{array}} \right] \text{.}
\end{equation}

Based on statistical considerations~\cite{lee2009polarimetric}, the mean parameters of the dominant scattering mechanism can be extracted from the 3-by-3 coherency matrix as a mean unit target vector   given by:
\begin{equation}
{{\bf{u}}_0} = {e^{j\phi }}\left[ {\begin{array}{*{20}{c}}
{\cos \bar \alpha }\\
{\sin \bar \alpha \cos \bar \beta {e^{j\bar \delta }}}\\
{\sin \bar \alpha sin\bar \beta {e^{j\bar \gamma }}}
\end{array}} \right] \text{.}
\end{equation}

It has been shown that the mean parameter $\bar \alpha$, ranging from $0$ to $\pi /2$ rad, is roll-invariant and that it is the main parameter for identifying the dominant ground scattering mechanism, while $\bar \beta$, $\bar \delta$ and $\bar \gamma$ can be used to define the target polarization orientation angle and $\phi$ is physically equivalent to an absolute target phase. The best estimate of $\bar \alpha$ is:
\begin{equation}
	\bar \alpha  = \sum\limits_{k = 1}^3 {{P_k}{\alpha _k}} \text{,}
\end{equation}
with ${P_k}$ representing pseudo probabilities:
\begin{equation}
	{P_k} = \frac{{{\lambda _k}}}{{\sum\limits_{i = 1}^3 {{\lambda _i}} }} \text{,} \qquad \sum\limits_{k = 1}^3 {{P_k}}  = 1 \text{.}
\end{equation}
Based on the value of $\bar \alpha$, a characterization of  the scatterer physical properties can be inferred, in terms of single (surface), volume or double bounce scattering, respectively corresponding to low, average and high values of $\bar \alpha$.

The coherence matrix eigen-decomposition can also be used to estimate polarimetric entropy $H$, a measure of the degree of statistical disorder (randomness) of the scatterer:
\begin{equation}
	H =  - \sum\limits_{k = 1}^N {{P_k}{{\log }_N}\left( {{P_k}} \right)} \text{,}
\end{equation}
with $N = 3$ for a monostatic radar and $N = 4$ for a bistatic radar. 
Polarimetric entropy ranges from $0$ to $1$. Low values of this parameter indicate weakly depolarizing targets (point scatterers), while high values characterize depolarizing targets (mixture of point scatterers). In the limit case of $H=1$, the target scattering corresponds to a random noise process.

As a complementary information to the polarimetric entropy, the polarimetric anisotropy $A$ can be estimated from the eigenvalues of the coherence matrix:
\begin{equation}
	A = \frac{{{\lambda _2} - {\lambda _3}}}{{{\lambda _2} + {\lambda _3}}} \text{.}
\end{equation}
Polarimetric anisotropy ranges from $0$ to $1$ and provides a measure of the relative importance of the second and the third eigenvalues of the eigen-decomposition.
Both $H$ and $A$ are roll-invariant parameters, as the eigenvalues are rotationally invariant.

\section{System Architecture}
\label{sec:sys_architecture}
Four subsystems compose the architecture of the proposed CBIR system:  data ingestion,  indexing,  query processing  and interactive visualization. The system allows users to query large-scale archives of polarimetric data and retrieve image tiles with  scattering characteristics similar to the ones of a user-provided query one. A schematic representation of the system is reported in Fig.~\ref{fig:highLevArch}.

The  visualization subsystem provides the interactive environment by which the  user --- typically an image analyst --- can communicate with the retrieval system and submit a query image tile.
The query is unsupervisedly analyzed to extract one or more vectors of content-based features. The resulting query signature feature vectors are sent to the the query execution subsystem, where the images with signatures most similar to the query image are efficiently retrieved. 
Finally, the retrieval results are shown to the user again by the user interface module.

The current section provides a detailed description of the architecture of the proposed CBIR system for polarimetric SAR data products, and analyzes the inner workings of each of the modules.

\subsection{Data ingestion}
\label{sec:sys_arch_ingestion}
Data ingestion is the first back-end module of the retrieval system. It consists of a system that retrieves images from the web or from a local database, extracts information about their contents in an unsupervised manner and stores it for later indexing. 

Workloads in this stage are typically characterized by strong variability.
For example, workload peaks are expected in early ingestion phases of a new database containing a large number of products.
On the other hand, low demanding or idle states are also expected to happen in a full-time operating system.
In order to avoid fast resource saturation or under utilization, the system has to be scalable, i.e. it has to be able to sustain both increasing and decreasing workloads with adequate performance by adapting hardware resources~\cite{herbst13elasticity}.

To this end, cloud-based computing environments can be exploited for scraping and performing processing and feature extraction operations on data products. 
In particular, elastic versions of such services can be implemented by allowing automated load balancers to on-demand provision and deprovision resources in the cloud environment. 
A load balancer allows the computing capacities to adapt to current requirements, i.e. to the amount of data that has to be ingested and processed at a certain time. Its operation is based on a distributed queue system and on a machine instance manager.

A further concrete issue are the heavy file transfers involved in this step in  cases in which the storage and data analysis systems are far apart (as is of course the case of our prototype). This can easily bring to a fast saturation of networking resources, adding to the cost of ingestion.

\subsubsection{Crawling and elastic load balancing}

\begin{figure}[t]
	\centering
	\includegraphics[width=0.5\columnwidth]{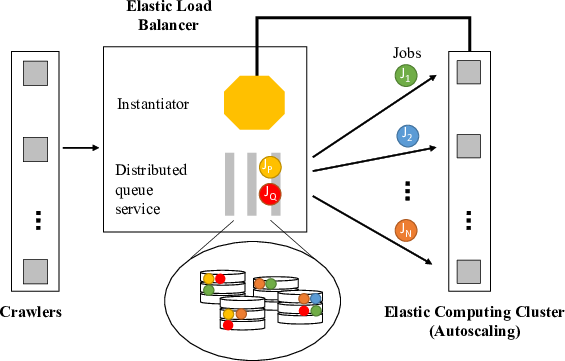}
	\caption{A schema of the elastic load balancing system. Crawling agents insert pointers to new products in a distributed queue service to mark them for ingestion; jobs are made available to machines in the elastic computing cloud, whose resources are managed by an instantiator module.}
	\label{fig:elasticLoadBalancer}
\end{figure}

Three components are involved in the first phase of the ingestion system: the crawling agents, the elastic load balancer, which is in turn formed by a distributed queue system and a machine instance manager, and the elastic computing cluster.
Crawling agents are charged with performing a continuous polling of the contents of the input data archive. In the specific case of the prototype under discussion, they perform an analysis of the NASA/JPL UAVSAR web pages with the aim of discovering new data products to ingest. 
When a new product is found, a pointer to it is created, and inserted in a distributed queue system. 
The distributed queue system is a scalable service that allows reliable queuing of independent jobs that cluster components have to perform. 
It acts as a buffer and communication channel among the crawling agents and the elastic cluster platform components, preventing the system to loose job instantiation requests when there is an excessive processing load or in case workers are scheduled to operate intermittently. 
The distributed nature of the queue also allows multiple interactions between crawlers and workers. 
In addition, a specific function of the queue manager is to allow better failure recovery strategies to be implemented in the system: when a worker node picks up a task to be executed, it also declares a timeout date after which it can be assumed that the processing will surely have been completed. 
If after this date the job has not been actively declared as completed by the worker tasked with executing it, the queue manager assumes that the processing failed and toggles a visibility flag on the scheduled job, which reappears in the queue for execution.
The other component of the load balancing system is the worker machine instance manager, which is in charge of managing the cluster capacities in terms of both the number and the kind of instantiated machines based on explicitly defined scaling policies. 
Scaling policies can be based on simple metrics like the queue size, feedbacks about resource usage from subsequent modules in the ingestion chain or on more complex strategies for costs or energy saving that could entail intermittent service availability. 

\subsubsection{Tiling and feature extraction}

\begin{figure}[t]
	\centering
	\includegraphics[width=0.5\columnwidth]{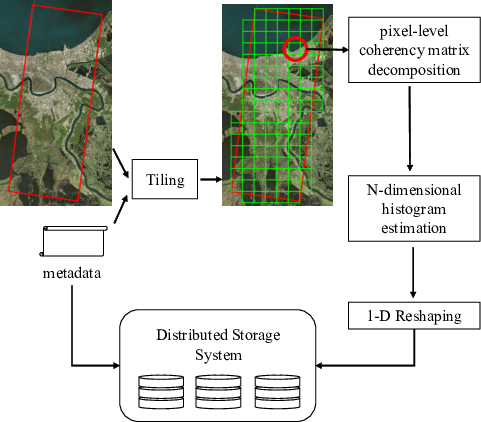}
	\caption{Block scheme of tiling and feature extraction.
	The specific feature extraction procedure shown refers to scattering matrix decompositions applied to polarimetric data products.}
	\label{fig:featureExtraction}
\end{figure}

Remote sensing image products cover vast areas of the Earth surface. Their content is therefore typically quite heterogeneous. Extracting a global description of each image product is not useful for most applications of content-based image retrieval systems, as, in general, the aim is to allow a user to identify specific targets or regions with similar characteristics.

The processing operations required for region-oriented content-based image mining consist of two main tasks: tiling the images in geo-referenced patches, whose size depends on the kind of sensor, on the resolution and on the aims of the content-based searches, and performing feature extraction operations.

The dimension of the tiles has to be chosen in such a way as to satisfy three main requirements:
\begin{itemize}
	\item Preserving as much as possible the local characteristics of each patch of the original image product: a tile should be small enough to make it probable that it will include only one target class.
	\item Retaining enough information for the extraction of the ground target category. This general requirement means, in the context of polarimetric SAR images, retaining a number of pixels that is reasonable for an accurate estimation of the $H/\alpha/A$ density distribution;
	\item Allowing efficient use of the addressing space. 
Because memory devices are arrays of bytes, for the sake of memory allocation efficiency the tile dimension should be chosen following a power-of-2 scheme.
\end{itemize}
Furthermore, an overlap between the tiles can also be accounted for.
During tile creations, image patches are univocally identified, georeferenced and processed for feature extraction. 
The block scheme of the operations implemented for polarimetric image ingestion is shown in Fig.~\ref{fig:featureExtraction}.

\subsubsection{Distributed storage}

The operations in the first phase are typically disk I/O bound and need to operate on data products that have volumes of Gigabytes (around 10 Gigabytes of average size for UAVSAR ground-projected products). 
The obtained results need to be accessed globally for index formation. 
This naturally raises the problem of allowing distributed access to files in clustering environments.
For this reason, the extracted image tile descriptors and metadata are pushed to a distributed file system.

A distributed file system allows storing information on networks of worker machines, offering at the same time scalability, fault tolerance and efficient computation performances.
The distributed nature of the systems allows large datasets to be accommodated on machines in the cluster.
Data descriptors are stored in files that are split into one or more equal and fixed size chunks that are replicated and stored across the network nodes.
Such operations allow the system to handle node failures, thereby preserving data from being lost and preventing the completion of long running computations on the cluster from depending on each single machine. 
Furthermore, distributed storage systems are designed to minimize the risk of network congestion and to increase overall system throughput.
As network resources have bandwidth limits, data movements have to be discouraged, preferring a paradigm that allows bringing computation close to data and not vice-versa.
Distributed file systems naturally fit such computational paradigms, as nodes working as storage units can also be employed as computational units in the cluster. 
We adopted for our system the Hadoop Distributed File System (HDFS)~\cite{shvachko2010hadoop}.



\subsection{Retrieval and indexing}
\label{sec:sys_arch_indexing}
\label{sec:TSVQ}

\begin{figure*}[t]
		\centering
		{\begin{minipage}{0.8\columnwidth}
		\begin{algorithm}[H]
						\label{fig:treeGrowingPseudoCode}
		\caption{Index generation}
    \begin{algorithmic}[1]
        \Function{index}{data}
            \State $model() \gets cluster(data)$													\Comment{Initialize model (root node)}
						\State $data \gets model(data)$ 															\Comment{Label the dataset}
            \State $leaves \gets find\_leaves(model())$										\Comment{Get leaves list}
						\While{$\left\vert{leaves}\right\vert \ne 0$}									\Comment{Iterate for each leaf}
								\State $leaf_{curr} \gets take(leaves)$
						    \If{not $stop criteria$}
							      \State $model() \gets cluster(data, model(), leaf_{curr})$
									  \State $data \gets model(data)$
										\State $leaves \gets find\_leaves(model())$
								\Else
										\State $leaves  \gets del\_leaf(leaves, leaf_{curr})$ \Comment{Prevent further bloomings}
								\EndIf
								
						\EndWhile
        \EndFunction
    \end{algorithmic}

		\end{algorithm}

		\end{minipage}}
		\caption{Pseudo code of the tree growing procedure used by the indexing subsystem.}
\end{figure*}

A crucial component for a large-scale CBIR system is an efficient  methodology for the generation of a global index. 
In the proposed tile-based retrieval system, the index is the result of an unsupervised hierarchical clustering subsystem that has the purpose to provide an efficient way to group image tiles that are characterized by similar scattering behavior and that are therefore all likely to be relevant to a specific query.
Without an index, the search engine would need to perform an $O(n)$ scan of the entire archive with unsustainable computational costs  for mining even moderate size archives.

In a ``big data" context, it is fundamental for the system to be able to perform indexing on large-scale datasets containing billions of entries, each representing the image content and eventually consisting of a large number of descriptors, and to allow fast and accurate retrieval results.
In order to generate the content-based index, the system has to analyze the set of tile descriptors generated by the feature extraction module.
Such analysis typically requires the application of machine learning algorithms.
However, indexing large scale archives of data requires suitable implementations, able to distribute the computation among multiple processors in the cluster. 
To this purpose, we propose a mechanism for scalable indexing based on Tree-Structured Vector Quantization~\cite{gersho1992vector}. 


\subsubsection{Tree Structured Vector Quantization}


\begin{figure}[t]
	\centering
	\includegraphics[width=0.5\columnwidth]{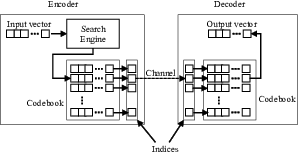}
	\caption{Encoding/decoding scheme in Vector Quantization. The encoder maps input vectors to a finite set of codewords and sends the codeword index through the channel. The decoder holds a lookup table and is able to find the original codeword based on the received index.}
	\label{fig:vectorQuantization}
\end{figure}

Vector Quantization is a signal processing technique that is popular in data compression. 
A generic schema of a vector quantizer consists of an encoding/decoding system.
Let us represent the generic entry (a set of tile descriptors, in our case) in the database ${\bf{D}}$ by a $d$-dimensional feature vector ${{\bf{x}}} = \left[ {{x_{1}},{x_{2}},...,{x_{d}}} \right]$.
The encoder maps the $d$-dimensional vectors in the vector space $R^d$ to a finite set of vectors $y_1, y_2, \dots, y_k$, called code vectors or codewords, operating according to a nearest neighbor or minimum distortion rule. 
We here consider the squared error distortion, i.e. the  square of the Euclidean distance between the  input vector and the codeword:
\begin{equation}
	d({\bf{x}},{{\bf{y}}_i}) = \left\| {{\bf{x}} - {{\bf{y}}_i}} \right\|_2^2 = \sum\limits_{j = 1}^d {{{({{x}}_j - {{{y}}_{i,j}})}^2}} \text{.}
\end{equation}
In this way, each codeword $y_i$ has an associated nearest neighbor region, also called a Voronoi region, defined by:
\begin{equation}
	{V_i} = \{ {\bf{x}} \in \mathbb{R}{^d}: \| {{\bf{x}} - {{\bf{y}}_i}} \| \le \| {{\bf{x}} - {{\bf{y}}_j}} \|, \; \forall j \ne i\} \text{,}
\end{equation}
such that
\begin{equation}
	\bigcup\limits_{i = 1}^k {{V_i}}  = {\mathbb{R}^d}, \qquad \forall i \ne j
\end{equation}
and
\begin{equation}
	\bigcap\limits_{i = 1}^k {{V_i}}  = \emptyset , \qquad \forall i \ne j \text{.}
\end{equation}
Once a codeword is associated to an input $d$-dimensional vector, the corresponding codeword index is sent to the decoding system through a channel (a file system or a communication link, depending on the application).
The decoder consists of a lookup table, i.e. a codebook containing all the possible codewords. When the index of a codeword is received, it will return the codeword corresponding to that index.
The schema of the encoding/decoding system is shown in Fig.~\ref{fig:vectorQuantization}.


Several different approaches can be considered in the  design a Vector Quantizer.
Tree-Structured Vector Quantization (TSVQ) is a class of constrained structure quantizers~\cite{cosman1993tree}.
In TSVQ, the codebook is constrained to have a tree structure.
The encoder builds the codeword associated to an input vector by performing a sequence of binary comparisons, following a minimum distortion rule at each branch, until a leaf (terminal) node is reached.
The path followed to reach the leaf node starting from the root indicates the binary sequence associated with the codeword.



A pseudo-code of the implementation of the tree-growing procedure is reported in~\ref{fig:treeGrowingPseudoCode}.
The tree structure is built starting from the root node. 
The root node is represented by the centroid of the entire set of feature vectors. 
From the root node, two new child nodes are estimated by applying the \mbox{$k$-means} clustering algorithm, each corresponding to the centroid of the space partitions ${\bf{D}} = \left\{ {{D_1},{D_2}} \right\}$ minimizing the within-cluster sum of squares cost function:
\begin{equation}
	{J_{WCSS}} = \sum\limits_{i = 1}^2 {\sum\limits_{x \in {D_i}}^{} {\left\| {{\bf{x}} - {{\bf{y }}_i}} \right\|_2^2} } \text{,}
\end{equation}
where ${{\bf{y}}_i}$ represents the i\emph{th} partition centroid. 
Then, each data point is assigned to its respective centroid by a binary string labeling, so that each group of data points defines a sub-tree.
The process continues iterating on the discovered nodes, until a predefined stopping criterion is reached.
The centroids of the items belonging to the leaves of the final tree structure, each with an associated binary string, represent the entries of the index.
During the growing process, each split produces a decrease of the average distortion and an increase in the average length of the binary string.
Based on these observations, several conditions are possible as stopping criteria.
For example, the algorithm can stop growing a sub-tree (thereby defining a leaf node) when one of the following events occurs:
\begin{itemize}
	\item the node contains a predefined minimum number of data points ${N_{\min }}$ under which further partitions become untrustworthy;
	\item the distortion measure given by the within-cluster sum of squares of the data vectors associated with the node is below a minimum threshold $WCSS_{\min }$;
	\item the maximum height of the tree ${h_{\max }}$ or, equivalently, the maximum binary code length have been reached.
\end{itemize}

\subsubsection{Scalable in-memory clustering}
\label{sec:inmem_clustering}

\begin{figure*}[t]
	\centering
	\includegraphics[width=0.7\columnwidth]{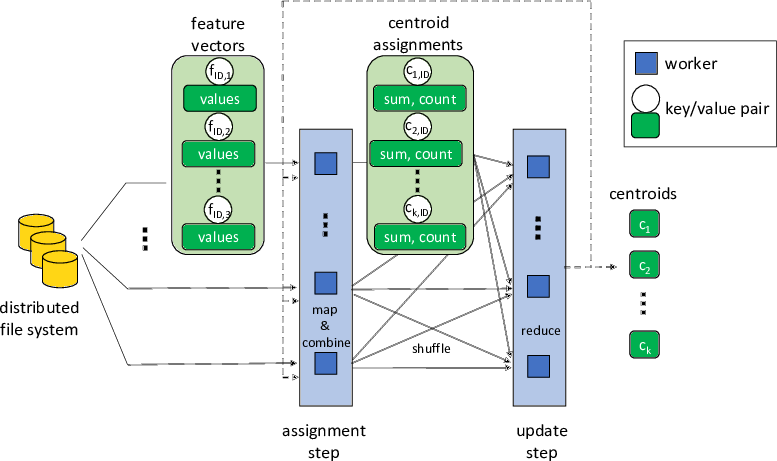}
	\caption{General MapReduce implementation of the \mbox{$k$-means} algorithm. The assignment and the update step can be implemented by a sequence of map\&combine and reduce phases. Tile descriptors stored in a distributed storage system are given as input to mappers together with the current centroids estimates. Each mapper performs the assignment step, putting out a key/value pair for each input vector, with the value being the input vector and the key being its nearest centroid. A combiner performs then a group-by-key, coupling to each centroid-key the sum and the number of its associated vectors. Reducers finally aggregate the sets of key/value pair to provide updated estimates of centroids.}
	\label{fig:mapRedClust}
\end{figure*}

The proposed indexing mechanism is based on a scalable version of the TSVQ algorithm. 
The method builds on a distributed implementation of the \mbox{$k$-means} algorithm to estimate the optimal space partitioning, allowing us to parallelize the clustering operations among the worker nodes of the computing cluster. 

A schema of a possible MapReduce implementation of the \mbox{$k$-means} algorithm is shown in Fig.~\ref{fig:mapRedClust}. 
Input data points are stored on a distributed file system as key/value pairs. 
The mapping phase consists of the so called \emph{assignment} step. 
At the i\emph{th} iteration, the set of $k$ centroid estimates ${{\bf{y}}_1}, \ldots ,{{\bf{y}}_k}$ is globally broadcast to map workers. 
Each map function has in input a group of data points and iteratively evaluates the squared Euclidean distance between the data points and each of the centroids vectors. 
The outputs consist of a key/value pair for each data point, with the value being the input vector and the key being its nearest centroid. 
Each map worker can perform a combine operation, coupling to each centroid key the sum and the number of its associated vectors.
This last local combination operation is in principle optional, yet it allows to avoid using expensive network resources by compressing the information generated by each node.
The \emph{update} step of the \mbox{$k$-means} algorithm is finally implemented by the reduce workers. 
Each reduce function takes in input the grouped-by-key intermediate key/value pairs and computes for each centroid the average of the associated values
\begin{equation}
	{\bf{y}}_i^{(t + 1)} = \frac{1}{{|D_i^{(t)}|}}\sum\limits_{{x_j} \in D_i^{(t)}} {{{\bf{x}}_j}}  \text{,}
\end{equation}
where $t$ is the current iteration and $D_i$ is the set of vectors belonging to centroid ${\bf{y}}_i$.
Each average value represents an updated centroid to be used in successive iterations. 
The algorithm stops when the update variations are below a given threshold.

The assignment step of the algorithm requires to repeatedly access data points and to evaluate the distances between each value vector and the cluster centroids:
\begin{equation}
	\label{eq:inMemoryEuclidean}
	\begin{aligned}
d({\bf{x}},{\bf{y}}_i^{\left( t \right)}) &= \| {{\bf{x}} - {\bf{y}}_i^{\left( t \right)}} \|_2^2 \\
																						 &= \| {\bf{x}} \|_2^2 - 2{\bf{x}} \cdot {\bf{y}}_i^{(t)} + \| {{\bf{y}}_i^{\left( t \right)}} \|_2^2 \text{.}
	\end{aligned}
\end{equation}
As the data points are fixed during iterations, the first term in the second line of the expression in Eq.~\ref{eq:inMemoryEuclidean} does not need to be recomputed at each iteration of the algorithm. 
Furthermore, also the third term, i.e. the centroid squared Euclidean norm, is repeatedly used inside each single iteration. 
This indicates the advantage of performing operations in-memory and avoiding, at each iteration, costly operations like both recomputing invariable quantities and storing and reading intermediate results on low performing devices. 
While standard cluster computing frameworks like Hadoop~\cite{white2009hadoop} routinely materialize to the distributed file system intermediate results produced by the directed acyclic graph of operations, with massive usage of cluster resources especially in the case of iterative processing schemes, in-memory ones like Apache Spark~\cite{zaharia2010spark, zaharia2012resilient} avoid this step and have therefore been selected as the base for our implementation.

Further improvements can be be obtained accounting for the sparsity of the feature vectors.
If the vectors are sparse, the computational complexity of performing \mbox{$k$-means} decreases from $O(ndki)$ to $O((n_{nz}+n)ki)$, where $n$ is the number of data points, $d$ is the data space dimensionality, $k$ is the number of clusters, $i$ is the number of iterations needed for convergence and $n_{nz}$ is the (average) number of non-zero elements in each data point.

\subsubsection{TSVQ Complexity}
\label{sec:TSVQComplexity}
The implemented TSVQ for scalable indexing processing works with a fixed number of clusters, $k=2$. This choice allows the association of binary codes to the leaves of the tree, each with a length equal to the level of the corresponding leaf. 
As each leaf corresponds to a partition centroid, all the points will be labeled with the binary code of the leaf they belong to.
Given the tree structure generated by the TSVQ algorithm and a query vector, retrieving the nearest feature vectors consists of traversing the tree from the root node and choosing, at each branch, the nearest child node, until a leaf is reached. The retrieved data points are the ones with the same codeword as the final leaf.

There are two main motivations behind the choice of the TSVQ algorithm for the index formation. First, this approach leads to the definition of a binary indexed tree, a structure that allows efficient lookup and update operations. 
In particular, search and modification operations can be executed in constant or $O(\log n)$ logarithmic average times, instead of the $O(n)$ linear access times required by linked lists. 
Second, as described in Sec~\ref{sec:inmem_clustering}, the \mbox{$k$-means} algorithm can be reimplemented to scale to massive sets of data.
In addition, scalable implementations of seeding strategies have been recently proposed, allowing for faster runs and more robust searches with respect to suboptimal solutions~\cite{bahmani2012scalable}. 


\subsection{Search and interactive visualization subsystems}

The interactive visualization subsystem provides users with a query interface to the search engine where they can select or upload a query/example image tile and set metadata constraints for the search (geographic coverage,  time of acquisition, sensor parameters, mission annotations).
The search subsystem is then invoked 
to transform the user provided examples to a form suitable for query processing, i.e. performing any pre-processing and feature extraction operations on them, and then to determine the best matching tiles based on the content index. 
Finally, the interactive visualization subsystem receives the identifiers of the best matching tiles and displays the results.



\section{Experiments}
\label{sec:experiments}


We considered full polarimetric SAR data products from NASA JPL's UAVSAR image archives. In the public archive, six images for each product are made available, that is the cross-products of the 4 Single Look Complex files representing the measurements of the scattering matrix $S_{hh}$, $S_{hv}$, $S_{vh}$ and $S_{vv}$. 
Each product is available in the ground-projected polarimetric format (equiangular geographic projection, \mbox{6-by-6 m} pixels resolution). 
In the experiments, data ingestion has been carried out by instantiating five nodes on the elastic cloud computing infrastructure by using the Amazon EC2 service\footnote{http://aws.amazon.com/ec2/}.
In the tiling phase, each image is divided in \mbox{512-by-512 pixel} sized square patches, corresponding to approximately \mbox{3-by-3 km} areas.
Then, for each tile, the feature extraction module performs pixel-by-pixel decomposition procedures, forms multi-dimensional histograms and pushes the vectorized tile descriptors to a durable cloud-based infrastructure, the Amazon S3 service~\footnote{http://aws.amazon.com/s3/}, in which data are redundantly stored across multiple facilities.
The reason we use such a durable storage service is that the results of the ingestion need to be reused in multiple experiments.
In this work, we performed the ingestion of 43 UAVSAR products, amounting to about 0.67 Terabytes of data, from which we obtained 12950 tiles, with a total size for the descriptors of 2.4 GB.
The replication factor in the HDFS filesystem has been set to three in all experiments.

\subsection{Scalability analysis}

\begin{table*}[t] 
\caption{Cluster worker node characteristics.
}
\centering
	\scalebox{1.00}{
  \begin{tabular}{lllllll}
			\toprule
        \multirow{2}{*}{ID} & \multicolumn{3}{c}{Processor}	& \multirow{2}{*}{vCPU}   & \multirow{2}{*}{Memory (Gb)}	& \multirow{2}{*}{Network bandwidth (Mbps)} \\
				\cline{2-4}
									& Model												 & Frequency (GHz) & Cache size (MB) 		& 	&		&  \\
			\midrule
			m1.small	  &  E5-2651 v2  & 1.80 & 30 		& 1				&1.70				& $\sim 300$~~~(Low) 				\\
			m1.medium		&  E5-2650 		 & 2.00 & 20 		& 1				&3.75				& $\sim 850$~~~(Moderate) 	\\
			m1.large 		&  E5-2651 v2  & 1.80 & 30 		& 2				&7.50				& $\sim 850$~~~(Moderate) 	\\
			m1.xlarge 	&  E5-2651 v2  & 1.80 & 30 		& 4				&15.00			& $\sim 1000$~(High	)				\\
			c1.medium		&  E5-2651 v2  & 1.80 & 30 		& 2				&1.70				& $\sim 850$~~~(Moderate)		\\
			c1.xlarge		&  E5-2650 		 & 2.00 & 20 		& 8				&7.00				& $\sim 1000$~(High	)				\\
	\bottomrule			
  \end{tabular}
	}
	\label{tab:amazonMachines}
\end{table*}

\begin{figure}[t]
	\centering
	\includegraphics[width=0.5\columnwidth]{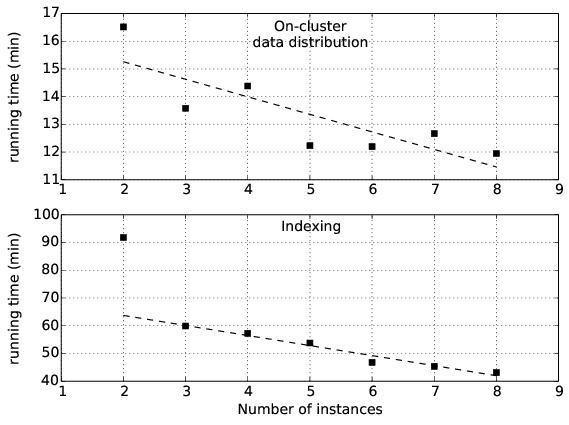}
	\caption{Horizontal scalability analysis. Black squares are running times in minutes by varying the number of instances in the cluster, while dashed lines are linear regression tests obtained by discarding possible outliers as determined by Bonferroni test ($\alpha<0.05$). The cluster instances are of type \emph{m1.medium} (see Tab.~\ref{tab:amazonMachines}). In (a) the times needed for the data transfer from the durable storage (S3 service) to the cluster distributed file system are shown, while (b) refers to the times needed for the indexing procedure. }
\label{fig:scalability}
\end{figure}

\begin{figure}[t]
	\centering
	\includegraphics[width=0.4\columnwidth]{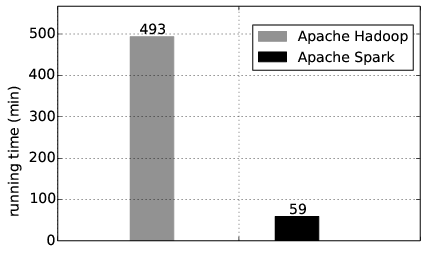}
    \caption{Running times in minutes of the indexing procedure obtained by Apache Hadoop, where performances are bounded by replication, serialization and disk I/O procedures needed for data sharing across parallel jobs, and Apache Spark, that is based on a shared memory-based architecture. In the experiments, the cluster is composed by 3 nodes and the database contains 12950 tiles.  }
\label{fig:scalability3}
\end{figure}

An experimental analysis of both vertical and horizontal scalability has been conducted by varying the cluster configuration.
Horizontal scalability is the ability to face increases in workload by adding worker entities to an application environment.
Fig.~\ref{fig:scalability} shows the results of the tests carried out by varying the number of nodes in the cluster.
The cluster instances are of type \emph{m1.medium} (see Tab.~\ref{tab:amazonMachines}).
The relation between the number of nodes and the times needed for data processing can be expected to be linear.
For this reason a least squares linear fitting has been carried out on the results and the presence of possible outliers has been assessed by Bonferroni method with familywise significance threshold $\alpha=0.05$.
The first plot refers to the time needed to transfer data from the durable (S3) storage to the cluster distributed storage (HDFS).
Since both storage systems are distributed, data transfer can be carried out in parallel and the linear relation is quite well satisfied.
The second plot refers to the indexing running times.
In this case, an outlier is detected for the simplest cluster, with two nodes, while the linear dependence is well satisfied for all larger configurations.
The reason for which in the two-nodes configuration the performance is worse and falls outside the linear dependence relation can be attributed to an overall shortage of memory.
This causes the loss of already computed in-memory intermediate results, thereby forcing their recomputation from HDFS and degrading the overall system performance.

The advantage of performing in-memory computations 
during the indexing procedure is better exemplified in Fig.~\ref{fig:scalability3}.
Standard frameworks for data analytics like Apache Hadoop would require a write operation at each iteration in order to save intermediate results on the distributed file system.
This operation involves overhead, as data must be distributed and replicated across the nodes, thereby increasing the total network traffic and slowing the computation further.
The results in Fig.~\ref{fig:scalability3} refer to an experiment performed with a cluster of 3 \emph{m1.medium} nodes.
In this case, exploiting in-memory computations improves indexing performance by almost 90\%.

\begin{figure}[t]
	\centering
	\includegraphics[width=0.5\columnwidth]{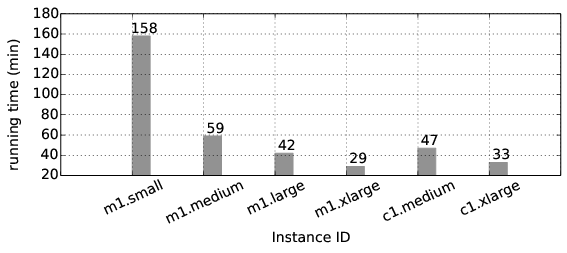}
    \caption{Vertical scalability analysis. Running times in minutes of the indexing procedure by varying the kind of instances in the cluster (see Tab.~\ref{tab:amazonMachines}). In the experiments, the cluster is composed by 3 nodes.}
\label{fig:scalability2}
\end{figure}

Vertical scalability is the ability to increase the computation capabilities of the cluster by improving the capacity of existing hardware or software.
In theory, the more the resources -- as for instance memory, network bandwidth, number of virtual CPUs -- we allocate in the cluster, the more the work that can be carried out in a certain amount of time. 
However, increasing the hardware capacities works just up to a certain point, as the best configuration depends on the algorithm, on the user needs and on their means.
In Fig.~\ref{fig:scalability2} we show an analysis of the indexing running times by varying the kind of instances allocated in the cluster, as reported in Tab.~\ref{tab:amazonMachines}.
The dimensionality of the cluster is kept fixed to three.
A first observation in this experiments is about the cheapest kind of instance, the \emph{m1.small}, which performs considerably worse than all other cases.
The reason is here a combination of scarceness of both memory and network resources.
On the other hand, the improvements in the other experiments are mainly attributable to the increase in the number of virtual CPUs allocated.

As for the system response time to queries, it is in the order of seconds and essentially independent of variations of the cluster settings.

\subsection{Cost analysis}

\begin{figure}[t]
	\centering
	\subfloat[]{\label{fig:costAnH} \includegraphics[width=6.2cm]{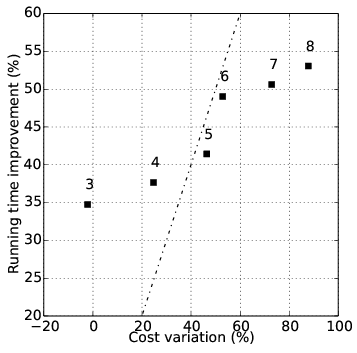}}
	\subfloat[]{\label{fig:costAnV} \includegraphics[width=6.1cm]{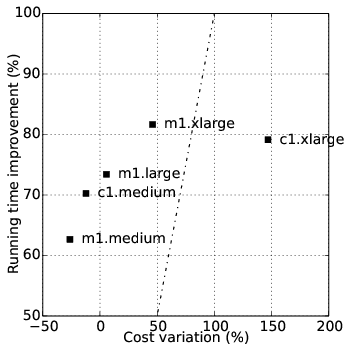}}
    \caption{Cost-benefit analysis. Improvement percentage of the indexing running time versus the cost variation percentage obtained by varying the cluster dimensionality (with \emph{m1.medium} instances), in (a), and the kind of instance used (with three nodes), in (b). The comparison is made with respect to the worst case, i.e. to a cluster composed by two nodes for (a) and to a cluster with \emph{m1.small} instances for (b). The straight dashed line in both figures is a line with unitary slope, representing the locus of the points where time improvements correspond to equal cost increments.}
\label{fig:costAnalysis}
\end{figure}

We also performed a cost-benefits analysis, in order to evaluate whether investing on more resources brings effective advantages, and to tune the system according to this analysis.
Results are reported in Fig.~\ref{fig:costAnalysis} and refer to the indexing phase.
The two plots show the relation between the improvement percentage of the indexing running time compared to the cost variation percentage obtained by varying the resources allocated in the cloud, with respect to the worst case, as described below.
The analysis for a variable number of nodes in the cluster is shown in Fig.~\ref{fig:costAnH}, where the instances are of type \emph{m1.medium} and the worst case for comparison is the use of two nodes (see Fig.~\ref{fig:scalability}).
This test shows that adding more nodes to the cluster diminishes the indexing running times, but increasing investments are not balanced by equivalent increases of performance.
The best trade-off has been obtained for the three nodes configuration for the specific case of the data volumes considered in the experiment.

Similar considerations hold for the analysis shown in Fig.~\ref{fig:costAnV}, which refers to the use of different types of instances, with a cluster dimensionality fixed to three and the worse case for comparison is the use of \emph{m1.small} instances (see Fig.~\ref{fig:scalability2}).
In this case, positive balances are obtained for both cases \emph{m1.medium} and \emph{c1.medium} instances, while the worse case happens for \emph{c1.xlarge} instances.

\subsection{Quality assessment}

\begin{figure}[t]
	\centering
	\includegraphics[width=0.4\columnwidth]{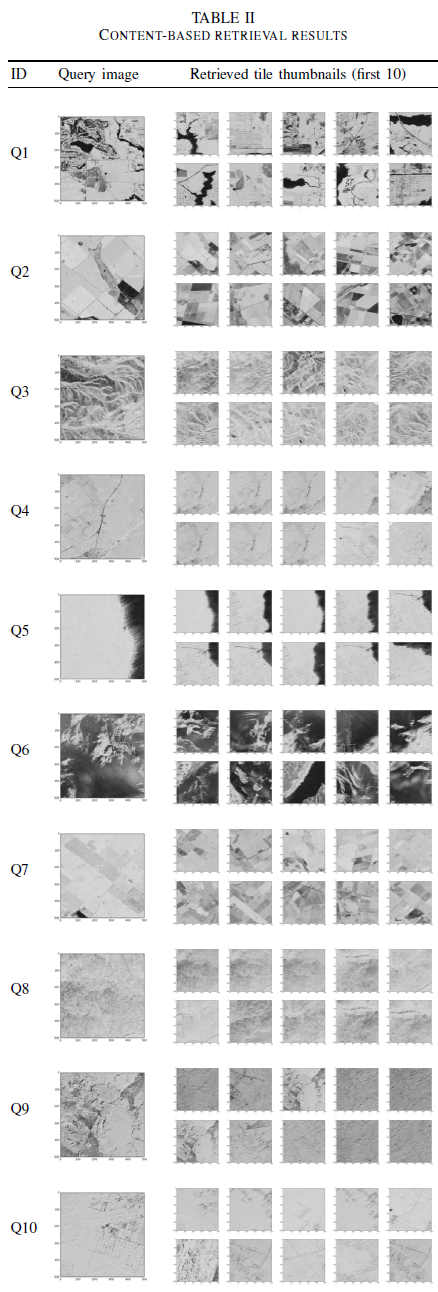}
\label{fig:sss}
\end{figure}

Evaluating the quality of content-based retrieval systems is a challenging task.
In the literature, quality assessment is not always present or is based on non-public databases of categorized images.
Evaluating the retrieval performance on massive archives of images is even more difficult.
A first idea to automatically assign labels to our dataset was to exploit publicly available worldwide land cover data, as those provided by ESA's Globcover project, obtained by MERIS sensor data.
UAVSAR products were coregistered with land cover maps and image patches were obtained from both sources.
Land cover tile descriptors consisted of the vectorized class labels and a separate index for such dataset was created.
We then tried to evaluate retrieval performance by querying the two search engines with coregistered example-query images, i.e. by polarimetric SAR and landcover images respectively.
However very poor results were obtained, due to differences in the spectral bands of the sensors ($15\mbox{--}30$ cm for L-band UAVSAR products and $390\mbox{--}1040$ nm for MERIS sensor), since the information collected about a target is different in the two cases, and to the variability in their spatial resolution (3~m and 300~m respectively).


Therefore we resort here to visual inspection analysis.
To this aim, the system has been repeatedly queried by different example-based queries.
The results of ten experiments, each characterized by different target characteristics, are shown in Tab.~II, where the query tile is shown together with quick looks of the top ten tiles displayed by the interactive visualization subsystem.
We consider that the retrieved tiles are highly consistent with the query images in terms of content similarity.
Better performances might be achieved by enriching the tiles content description by textural  and  shape  descriptors  as  well  as  more advanced  descriptions  for  metric  resolution  data  based  on signal  decomposition  in  fractional  frequency  domains as in~\cite{walessa2000model,singh2013sar}.


\section{Discussion and Conclusions}
\label{sec:conclusions}

Content-based EO image retrieval is an open research area that is currently receiving particular attention because of the fast and continuous increase in the volume of acquired product repositories. 
While several systems for image mining have been proposed, they often lack the capability of managing and accessing massive amounts of data because of their architectural and algorithmic approaches oriented to limited archive volumes.

We propose using inexpensive computing clusters instantiated on public Infrastructure-as-a-service cloud platforms in order to ingest and index Petabyte-scale remote sensing image archives with the aim of implementing query-by-example retrieval functionalities on them.
To this end, we developed and tested a full prototype operating on a subset of polarimetric SAR products from the NASA JPL UAVSAR archive.

The ingestion component of the system is centered on an elastic load balancer instantiating virtual resources in elastic platforms based on the occupation of input queues, in order to distribute ingestion processes on multiple machines. 
Furthermore, the indexing component is a parallel/scalable indexing algorithm built upon a cluster-computing framework to enable the prototype to face the difficulties of performing large scale processing on massive image archives.
The use of public Infrastructure-as-a-service cloud platforms has proved both flexible, by providing dynamic scaling of resources, and advantageous, as it allows economies of scale.
Performances of retrieval measured  have proved both the efficiency of the system in terms of response times and its effectiveness in terms of retrieval quality.
Although proposed for the analysis of polarimetric SAR image archives, the proposed approach is general and can be expanded to ingest archives from different sensors and missions.



\ifCLASSOPTIONcaptionsoff
  \newpage
\fi

%
%
\bibliographystyle{plain}
\bibliography{bibliography_bigdata}

%
%
%

\end{document}